\documentclass[prl, twocolumn, superscriptaddress, showpacs]{revtex4}  
\usepackage{amssymb}
\usepackage{amsmath}
\usepackage{bbm}
\usepackage{graphicx}

\def\Im{\hbox{Im}}

\begin{document}

\author{Philipp Werner}
\affiliation{Theoretische Physik, ETH Zurich, 8093 Z{\"u}rich, Switzerland}
\author{Emanuel Gull}
\affiliation{Department of Physics, Columbia University, 538 West, 120th Street, New York, NY 10027, USA}
\author{Olivier Parcollet}
\affiliation{Institut de Physique Th{\'e}orique, CEA, IPhT, CNRS, URA 2306, F-91191 Gif-sur-Yvette, France}
\author{Andrew J. Millis}
\affiliation{Department of Physics, Columbia University, 538 West, 120th Street, New York, NY 10027, USA}

\title{Two-stage metal-insulator transition in the 2D Hubbard model: momentum selectivity in the 8-site dynamical cluster approximation}

\date{March 15th,  2009}

\hyphenation{}

\begin{abstract}
Metal-insulator transitions
in the paramagnetic phase of the two dimensional square lattice Hubbard model 
are studied using the dynamical cluster approximation with eight momentum cells. We show that 
both the interaction-driven and the doping-driven transition are  multi-stage and momentum-sector specific, with Fermi liquid metal and fully gapped insulator phases separated by an intermediate 
phase in which some regions of the Brillouin zone are gapped while others sustain 
gapless quasiparticles.  We argue that this is the coarse-grained version of a gradually shrinking arc or pocket.  
A pronounced particle-hole asymmetry is found.
\end{abstract}


\maketitle

The metal-insulator transition in strongly interacting Fermion systems is one of the central  issues in condensed matter physics \cite{Imada98}.  The transition observed in the  high-$T_c$ cuprate family of materials is particularly interesting. As the doping $\delta$ (here defined as the difference of carrier concentration from 1 carrier per CuO$_2$ unit) is reduced to zero the materials become less metallic, with an insulating phase occurring at zero doping. Experiments,  including analyses of the temperature dependence of the London penetration depth \cite{Lee97,Millis98}, normal state  \cite{Tacon05} and superconducting \cite{Tacon06,Guyard08} Raman spectra,  $c$-axis conductivities \cite{Ioffe98}, photoemission \cite{Shih08} and scanning tunneling microscopy \cite{Kohsaka08} (for a review see \cite{Huefner08}) suggest that the transition is driven by a doping-dependent  decrease in the size of the region of momentum space which can support well-defined mobile quasiparticles. Understanding this behavior has proven theoretically challenging. Weak coupling renormalization group calculations \cite{Zanchi98,Laeuchli04} indicate  a parameter regime in which interactions 
involving electrons near the $(\pm \pi/2,\pm \pi/2)$ points flow to weak coupling while interactions important for electrons  near the $(\pi,0)$ and $(0,\pi)$ portions of the Fermi surface flow to strong coupling for reasons relating to antiferromagnetism \cite{Zanchi98} or Mott physics \cite{Laeuchli04}. However, the basic result is a flow to a strong coupling regime where the equations no longer apply and the physics must be studied by other methods.

\begin{figure}[tbh]
\begin{center}
\includegraphics[angle=0, width=0.43\columnwidth]{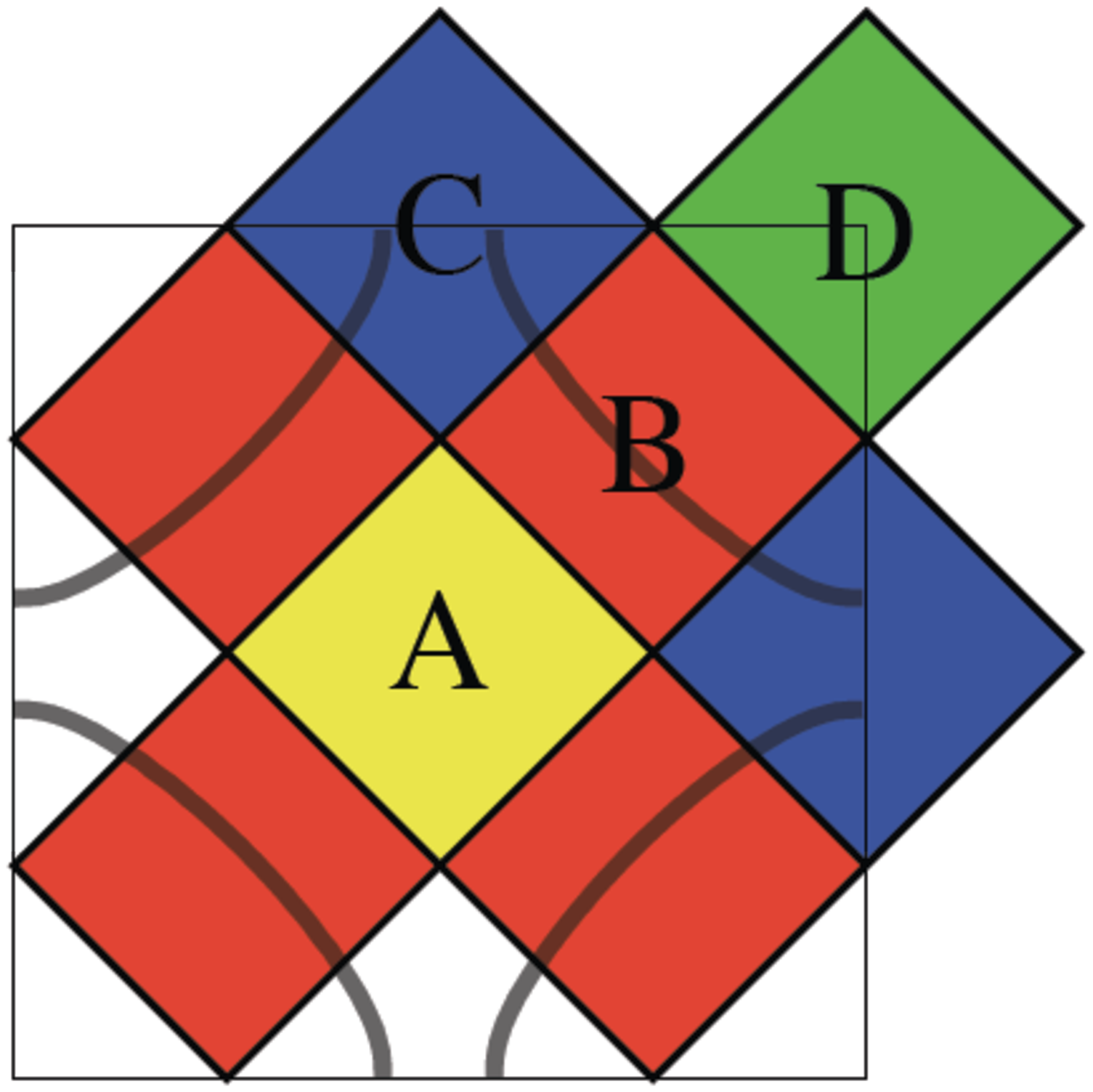}
\hfill
\includegraphics[angle=0, width=0.52\columnwidth]{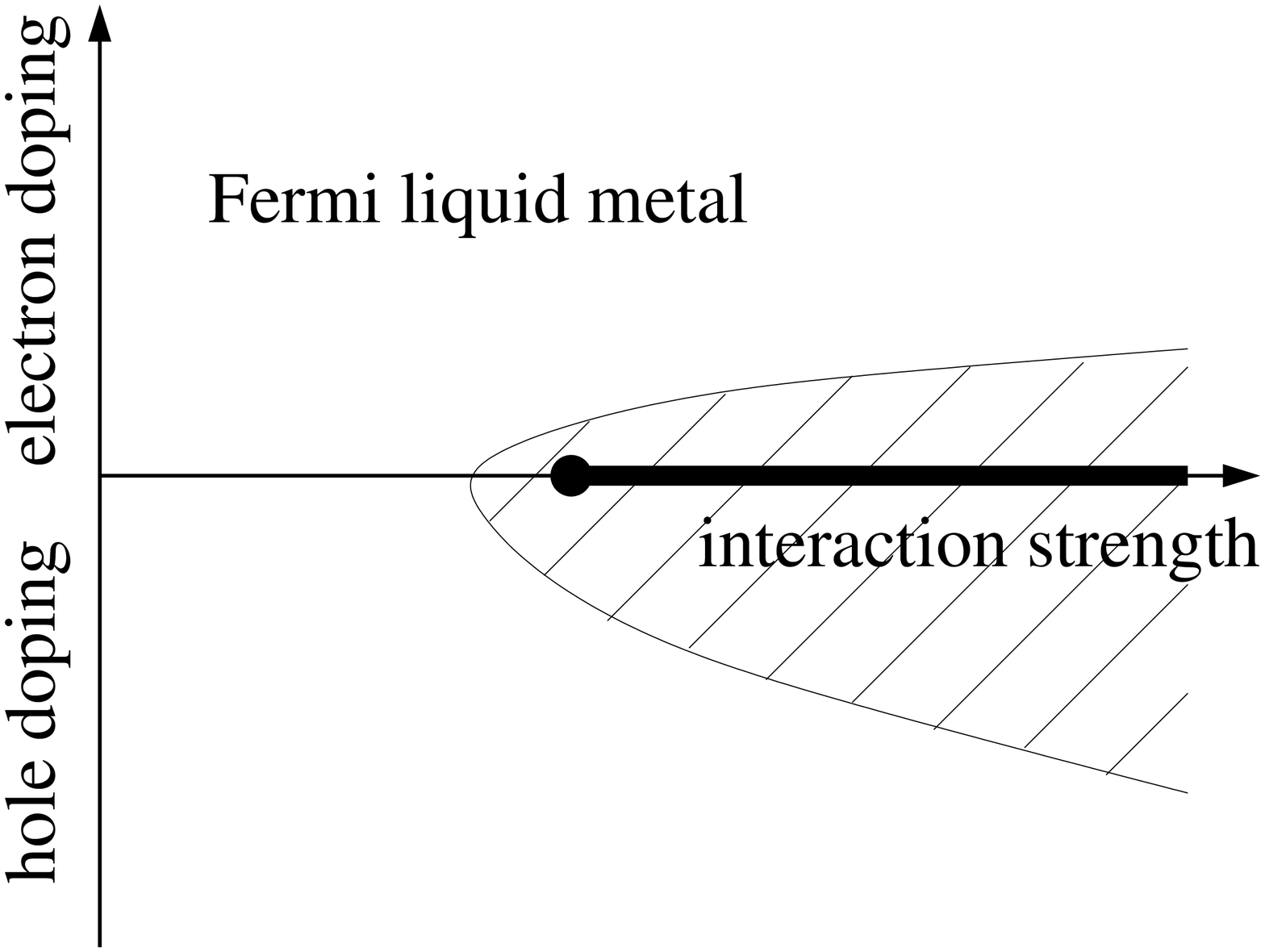}
\caption{Left panel: Brillouin zone partitioning associated with the 8-site cluster
and definition of the four inequivalent momentum sectors $A$, $B$, $C$  and $D$.
The noninteracting Fermi surface for $t'=-0.15t$ and density $n=1$ is indicated by the gray line.
Right panel:  sketch of the paramagnetic state DCA phase  diagram of the Hubbard model  on this cluster showing a Fermi liquid metal phase, an intermediate phase (hashed region) and the 
insulating state (heavy horizontal line). 
}
\label{cluster}
\end{center}
\end{figure}

Dynamical mean field theory (DMFT) \cite{Georges96,kotliar_review_rmp_2006} has provided valuable insights into the physics of strongly correlated materials. Its original single-site version averages physical quantities over the entire Brillouin zone. 
Cluster extensions of DMFT provide some momentum resolution \cite{Hettler98,Lichtenstein00,Kotliar01,Okamoto03,Maier05} and have indicated interesting momentum-dependent pseudogap behavior \cite{Parcollet04,Civelli05,Kyung06,Chakraborty07,Haule07,Park08,Gull08_plaquette,Ferrero08}. However, most of these works were based on studies of two and four-site clusters where 
the momentum resolution is so coarse that information about variation around the Fermi surface must be inferred from averages involving momentum regions far from the Fermi surface. Important exceptions are the work of Macridin {\it et al.}  and Vidhyadhiraja {\it et al.}  who studied 16 site clusters which have the momentum resolution needed to capture the variation around the Fermi surface directly, identifying a pseudogap phase at doping $0.05$ \cite{Macridin06} and a critical point defined from the self energy \cite{Vidhyadhiraja08}. We comment on the relation of our results to these works in the conclusion.

In this paper we use a recently developed continuous-time auxiliary field quantum Monte-Carlo algorithm [28] to perform direct computations of  the momentum variation of the electron spectral function and self  energy around the Fermi surface for the 8-site cluster shown in Fig.~\ref{cluster} and
for a range of interaction strengths and carrier concentrations.
An  attractive feature of this cluster is that  the Fermi surface of the non-interacting problem (shown as the heavy line) passes through two symmetry-independent sectors (labeled $B$ and $C$), so that a coarse-grained approximation to the variation around the Fermi surface can be studied directly. 
We find a two-stage metal insulator transition in both the interaction-driven and doping driven cases: the Fermi liquid metal and fully gapped insulator are separated by an intermediate phase, shown as the hashed region in the right panel of Fig.~\ref{cluster}, in which sector $C$ is gapped while sector $B$ remains gapless.  
The metal-insulator transition is a momentum-selective transition of the kind discussed by Ferrero {\it et al.} \cite{Ferrero08}. 

The one-band Hubbard model in two dimensions reads:
\begin{equation}
H = \sum_{p,\sigma} \epsilon_{p}  
c^\dagger_{p,\sigma}c_{p,\sigma}+U\sum_i n_{i,\uparrow}n_{i,\downarrow},
\end{equation} 
with $\epsilon_p=-2t(\cos(p_x)+\cos(p_y))-4t'\cos(p_x)\cos(p_y)$  denoting the electron dispersion and $U$ the on-site repulsion.
We are interested in the low-temperature behavior as a function of interaction strength $U$  and doping $\delta$.
We use the DCA formulation of cluster DMFT \cite{Hettler98,Maier05}, in which the Brillouin zone is divided  into  $N=8$  ``patches'' defined by the basis functions $\phi_\alpha(p)$, which are 1 for $p$ in patch $\alpha$ and zero otherwise. The momentum dependence of the self-energy becomes $\Sigma(p,\omega) \rightarrow \sum_{\alpha=1}^{8}\phi_\alpha(p)\Sigma_\alpha(\omega)$, and the frequency dependent functions $\Sigma_\alpha(\omega)$, $\alpha=1,\ldots,8$ are obtained from the 
solution of an appropriately defined  $8$-site quantum impurity model \cite{Maier05}.  We solve the impurity model with the continuous-time auxiliary field technique \cite{Gull08_ctaux} with delayed updates \cite{Alvarez08}. Results are presented for $t'=0$ (where the less severe sign problem allows us to access lower temperatures) and  $t'=-0.15 t$, which is computationally accessible but more ``generic" than $t'=0$, because the electron-hole symmetry is broken and the van Hove singularity is removed from the half-filled Fermi surface.  The lowest temperature accessible with the computational resources available to us is $\beta t\approx 40$ at $U < 8t$ corresponding (with the conventional high-$T_c$ band parametrization  \cite{Andersen95}) to $\sim 100K$.

We probe the transition via the sector-specific impurity model Green's functions evaluated at the mid-point $\beta/2$ of the imaginary time interval. This quantity is directly measured in our simulations and is related to  the 
value of the spectral function of the $\alpha$ sector $A_\alpha(\omega)$ near the Fermi level ($\omega=0$) 
by $\beta G_\alpha(\beta/2)=  - \int \frac{dy}{ \cosh(y)}A_\alpha(\frac{2y}{\beta})$. 
%
%
\begin{figure}[t]
\begin{center}
\includegraphics[angle=0, width=0.52\columnwidth]{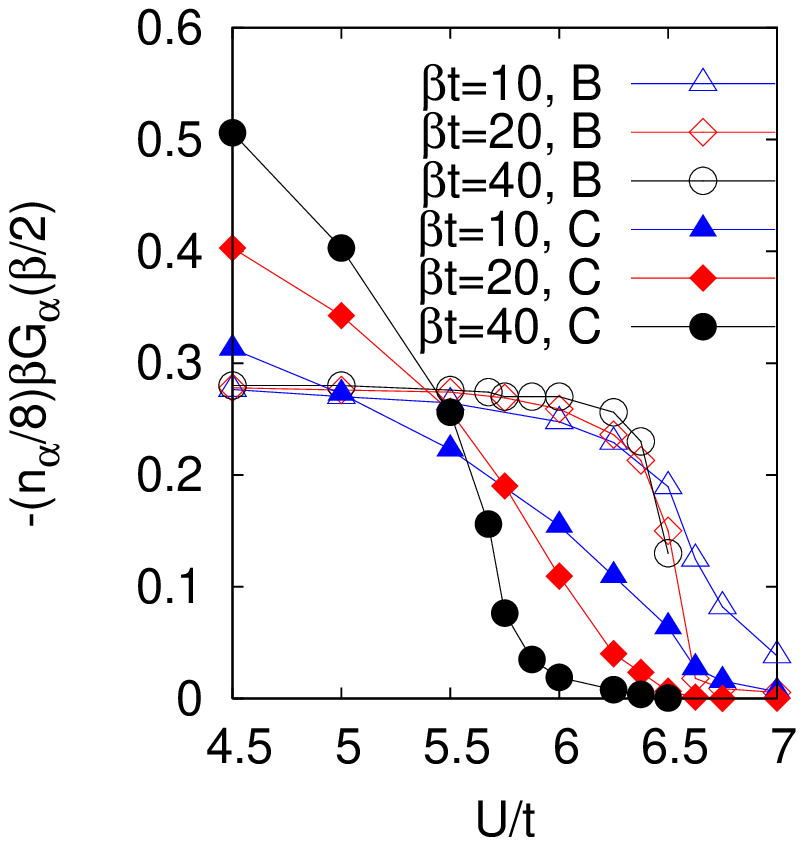}\hfill
\includegraphics[angle=0, width=0.479\columnwidth]{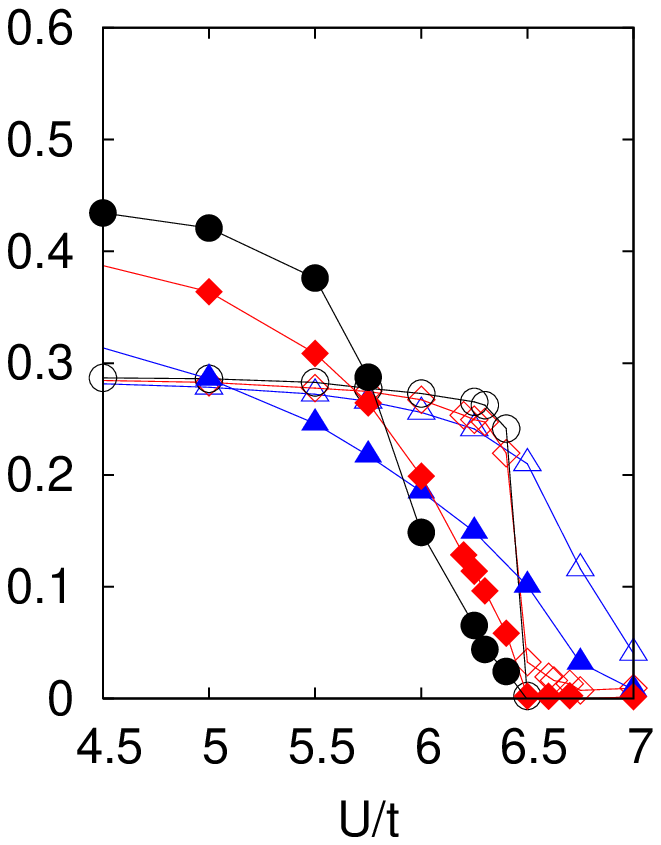}
\caption{
Sector $B$ (open symbols) and $C$ (filled symbols) Green's functions, scaled by the number of equivalent sectors $n_\alpha$ and evaluated at the midpoint of the imaginary time interval, as a function of $U$ at half filling. The left panel shows results
for $t'=0$; the right panel for $t'=-0.15t$.
}
\label{G_u}
\end{center}
\end{figure}
Figure~\ref{G_u} shows our results for the sectors ($B$ and $C$) which contain the Fermi surface at half filling. In sector $B$ (open symbols) we see for $U\lesssim 6.3t$ a weakly temperature dependent value which is consistent with the noninteracting Fermi-surface density of states at density $n=1$.  For $U\gtrsim 6.5t$ the density of states is very small, indicating a gapped state.  The non-interacting density of states in sector $C$ (full symbols) is larger and more strongly temperature dependent due to the van Hove singularity, which at $t'=0$ is at the  half filled Fermi level, and at $t'=-0.15t$ is about $0.2t$ below. 
We see that the temperature dependence changes sign  at  $U \approx 5.5t$ ($t'=0$) and  $U\approx 5.8t$ ($t'=-0.15t$); for larger $U$ the evolution is towards an insulating state. Thus sector $C$ undergoes a gap opening transition at a smaller $U$ than sector $B$, both for $t'=0$ and $t'=-0.15 t$. We emphasize that this two-stage transition is directly visible in our data and is a result enabled by the momentum resolution of the 8-site cluster. It is different from the result of single-site DMFT, which predicts a single transition at a large $U\sim 1.5W$ with $W$ (here $8t$) the full bandwidth and from the result of 4-site cluster calculations, which predict a single transition at $U\sim 4.2t$ (DCA \cite{Gull08_plaquette}) or $\sim 5.5t$ (CDMFT \cite{Park08}).

\begin{figure}[t]
\begin{center}
\includegraphics[angle=0, width=0.499\columnwidth]{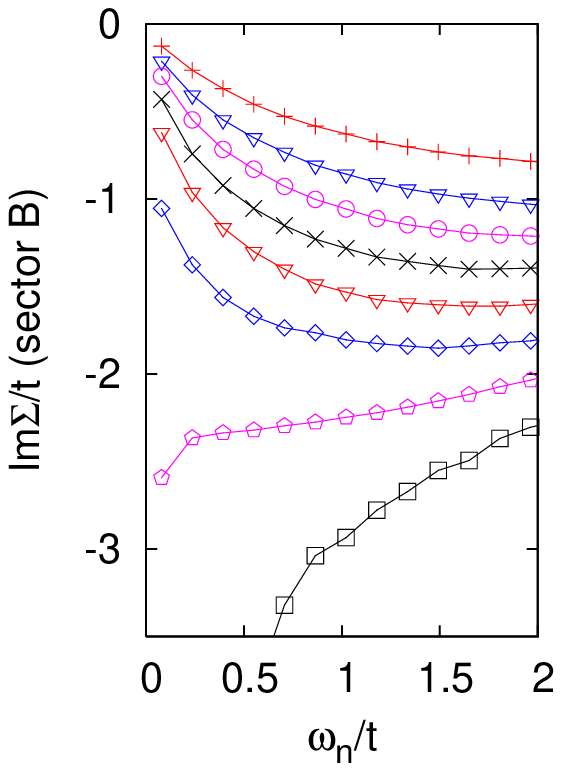}\hfill
\includegraphics[angle=0, width=0.499\columnwidth]{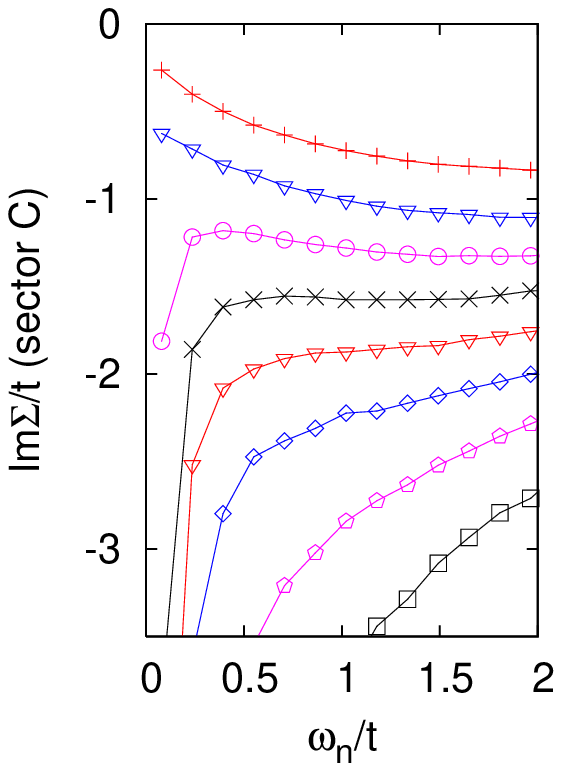}
\caption{Imaginary part of the self energy in momentum sectors $B$ and $C$ for half-filling and $t'=0$. The interaction strengths are (from top to bottom) $U/t=5$, $5.5$, $5.75$, $6$, $6.25$, $6.375$, $6.5$, and $6.625$. 
}
\label{sigmau_sector}
\end{center}
\end{figure}

Figure~\ref{sigmau_sector} displays  the imaginary part of the Matsubara axis self energy in sectors $B$ and $C$ for several $U$ in a range  containing the two transitions.   We present results obtained for $t'=0$  both because the absence of the sign problem allows us to obtain more accurate data and because the particle-hole symmetry implies Re$\Sigma=0$ which simplifies the discussion, but our results for $t'=-0.15t$ are consistent. The smaller $U$ curves of Fig.~\ref{sigmau_sector} show the low frequency behavior expected in a Fermi liquid phase $\text{Im} \Sigma(i\omega_n)\sim-\omega_n (Z^{-1}-1)$  with quasi-particle residue $Z$ decreasing as $U$ is increased towards the transition. The larger $U$ curves reveal the  low frequency behavior $\Im \Sigma(i\omega_n)\sim -\Delta^2/\omega_n$ expected in an insulating phase, with $\Delta$ decreasing as $U$ is decreased towards the metal insulator transition value.  In the intermediate regime $5.5\lesssim U/t \lesssim  6$, $\Sigma_C$ exhibits an insulating behavior while $\Sigma_B$ appears consistent with the Fermi liquid expectation. 

\begin{figure}[t]
\begin{center}
\includegraphics[angle=0, width=0.9\columnwidth]{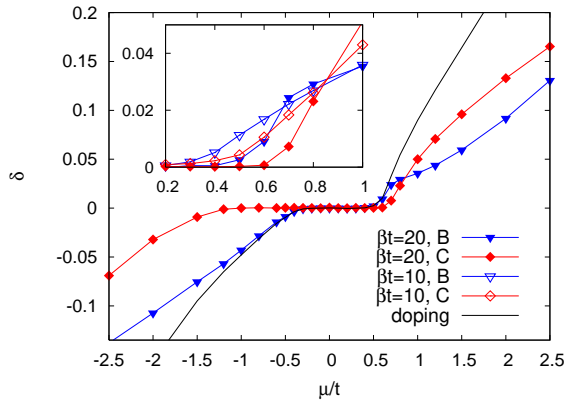}
\caption{
Main panel: $B$ and $C$ sector dopings (summed over spin and sector degeneracy) and total doping as a function of chemical potential for $U/t=7, t' = -0.15 t$ at $\beta t = 20$. Inset: expansion of electron-doped metal insulator transition region showing the temperature dependence. 
}
\label{n_u7}
\end{center}
\end{figure}

Figure~\ref{n_u7} presents the doping driven transitions for $U=7t$, large enough that at half filling all sectors are gapped.  We denote by $\mu$ the chemical potential minus the Hartree shift $Un/2$. For $t'=-0.15 t$ hole-doping leads to a wide range of chemical potentials where sector $C$ remains insulating while sector $B$ is doped.  On the electron-doped side, the two transitions occur almost simultaneously and the onset of doping in sector $B$ appears very sudden. 
However, the inset to Fig.~\ref{n_u7} shows that even on the electron-doped side there is a small window of chemical potential ($\mu \sim 0.7t$) where sector $C$ develops a gap but sector $B$ remains gapless. In the model with $t'=0$ (not shown) 
the $B$-sector transition (at $\beta t=40$) occurs near $|\mu|/t =0.5$ and the $C$-sector transition near $|\mu|/t=1$. 

\begin{figure}[t]
\begin{center}
\includegraphics[angle=-90, width=0.9\columnwidth]{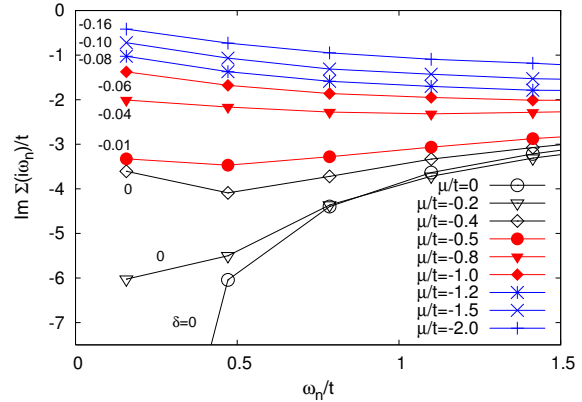}
\caption{
Imaginary part of $B$-sector self energies for the doped Hubbard model at $U=7t$, $t'=-0.15 t$ and $\beta t=20$. Traces indicated by open circles, triangles and diamonds correspond to $\mu$ where both sector $B$ and sector $C$ are gapped; traces marked by full circles, triangles and diamonds correspond to $\mu$ where sector $C$ remains undoped and sector $B$ is doped, while traces marked by asterisks, crosses and plus signs correspond to $\mu$ where both sectors are doped.
}
\label{sigma_u7_B_hole}
\end{center}
\end{figure}

The particle-hole asymmetry  in the model with $t'=-0.15 t$ is evident also in the self energies. We find that on the electron-doped side, at high dopings $\delta\gtrsim 0.15$ the two sectors have the same quasi-particle residue ($Z_{\alpha} \approx 0.5$ for $\delta = 0.23$);  as doping is decreased, $Z$ decreases in both sectors, but at a slightly faster rate in sector $C$.  Down to a doping of $\delta \approx 0.08$, $Z$  is reasonably well defined in both sectors, is larger  than $0.2$  and is only weakly anisotropic. For lower dopings in the vicinity of the $C$-sector gapping the behavior is not Fermi liquid like and the $C$-sector self energy becomes much larger. Figure~\ref{sigma_u7_B_hole} shows that the hole-doped side exhibits a  richer behavior. An analysis of the real and imaginary parts of $1/\Sigma$ (not shown) reveals that at $-0.4t<\mu$ where the model is insulating (gaps in all sectors) the low frequency self energy is dominated by a pole $\Sigma _{\alpha} \sim \Delta_{\alpha}^2/(i\omega_{n}-\omega_p^{\alpha})$ with pole strengths $\Delta_{B}/t\approx 3$  and $\Delta_{C}/t\approx 5$  and  position $\omega_p^{\alpha}$ shifting with chemical potential. This shift is responsible for the changes evident in the three undoped traces.  In the sector selective regime where $B$ is doped and $C$ is not, the $B$ self energy is  that of a `bad metal' ($\lim_{\omega_n\rightarrow 0} \text{Im} \Sigma(i\omega_n) \neq 0$): any  Fermi liquid coherence scale would be far below the temperature regime we can access. However, as the doping is increased into the regime where all sectors are ungapped the self energy evolves towards Fermi liquid behavior (with $Z_{\alpha}\approx 0.4$ for $\delta=-0.23$). 




The results are intriguingly similar to data on high-$T_c$ materials where a growing body of evidence indicates a metal-insulator transition in the hole doped materials apparently occurring by a reduction in the size of the $k$-space regions which can support mobile particles, while the electron-doped materials behave more conventionally (apart from the effects of long ranged commensurate density wave order not included here).   We believe that our coarse momentum resolution approximates the continuous reduction of Fermi surface area by a two step transition. Preliminary results for a 16-site cluster with Fermi surface crossing 4 inequivalent sectors indicate a 4 stage metal-insulator transition consistent with the gradual opening of a gap near $(0,\pi)$. The electron-hole asymmetry suggests an important role played by the van Hove singularity in the sector-selective transition.  On the electron-doped side the sector-selective region shrinks with increasing particle-hole asymmetry and may completely vanish at $t'=-0.3t$.


One can view the differentiation between sectors $B$ and $C$ as an orbitally selective transition of the effective impurity model \cite{Ferrero08}. 
It is known that orbitally selective transitions may be destabilized by a Kondo coupling to other sectors; however in our case the gapped sector $C$ is two-fold degenerate and we believe that the two electrons in this sector are strongly antiferromagnetically coupled (or form singlets) so that the state is stable against a Kondo coupling. We therefore expect that the effect survives beyond the DCA approximation we have used.

Our results are in general agreement with previous cluster DMFT work \cite{Civelli05,Macridin06} indicating a `pseudogap' phenomenon more pronounced for hole-doped than electron-doped materials. Our improved momentum resolution confirms that the effects inferred from two and four-site calculations  \cite{Parcollet04,Civelli05,Chakraborty07,Haule07,Kyung06, Park08,Gull08_plaquette} indeed reflect behavior associated with near-Fermi-surface states, while the improved temperature resolution and the examination of a range of interactions and dopings  reveals the sector-specific  metal-insulator transitions which underly  the pseudogap observed by Macridin {\it et al.} \cite{Macridin06}.  Note that while Ref.~\cite{Macridin06} reported a large density of states near $(0,\pi)$ for 5\% doping both for $t'=0$ and   $t'=-0.3t$, we find (for $t'=0$ and $t'=-0.15t$) that dopants first appear near the  {\it nodes} (sector $B$) with sector $C$ remaining gapped. However, after the initial doping, the sector $C$ occupancy increases rapidly, so the difference may be a doping-level effect.  The self energy transition studied in Ref.~\cite{Vidhyadhiraja08} can be seen in Fig.~\ref{sigma_u7_B_hole} and the corresponding data for sector $C$ (not shown)  as a change from $\text{Im} \Sigma(i\omega_n)\sim \omega_n$ to a flatter $\omega$-dependence as $\delta$ drops below  $\sim 0.15$. This change occurs well before the sector $C$ gapping transition $(\delta \sim 0.08$) and its connection to the sector-selective transitions we find remains to be clarified.

The important open questions concern the physical origin of the gapping behavior. We find cluster antiferromagnetic correlations which are large and grow rapidly in the vicinity of the metal-insulator transitions. However, in the model with $t'=-0.15t$ the correlation functions do
not change dramatically across the sector-selective transitions, suggesting that the phenomenon may not be antiferromagnetically driven, perhaps supporting the interpretation of L{\"a}uchli {\it et al.} \cite{Laeuchli04}. 
Optimized implementations of the hybridization expansion technique \cite{Werner06} should permit  examination of the cluster eigenstates along the lines of Ref.~\cite{Gull08_plaquette}, and thus provide direct information on the role of singlet formation and antiferromagnetism.

We thank A. Lichtenstein and T. M. Rice for helpful conversations  and acknowledge support from NSF-DMR-0797074 (AJM, partial support EG and PW),  the Swiss National Science Foundation (PP002-118866) and the ANR 
under grant ECCE. 
The calculations were done on the Brutus cluster at ETH Zurich using ALPS \cite{ALPS}.

\def \bibfnamefont#1{#1}
\def \bibnamefont#1{#1}

\end{document}